# Investigating the Change of Web Pages' Titles Over Time

Martin Klein
Department of Computer Science
Old Dominion University
Norfolk, VA, 23529
mklein@cs.odu.edu

Michael L. Nelson
Department of Computer Science
Old Dominion University
Norfolk, VA, 23529
mln@cs.odu.edu

## ABSTRACT
Inaccessible web pages are part of the browsing experience. The content of these pages however is often not completely lost but rather missing. Lexical signatures (LS) generated from the web pages' textual content have been shown to be suitable as search engine queries when trying to discover a (missing) web page. Since LSs are expensive to generate, we investigate the potential of web pages' titles as they are available at a lower cost. We present the results from studying the change of titles over time. We take titles from copies provided by the Internet Archive of randomly sampled web pages and show the frequency of change as well as the degree of change in terms of the Levenshtein score. We found very low frequencies of change and high Levenshtein scores indicating that titles, on average, change little from their original, first observed values (rooted comparison) and even less from the values of their previous observation (sliding).

## 1. INTRODUCTION

Inaccessible web pages and "404 Page Not Found" responses are part of the web browsing experience. Despite guidance for how to create "Cool URIs" that do not change [4] there are many reasons why URIs or even entire websites break [16]. However, we claim that information on the web is rarely completely lost, it is just missing. In whole or in part, content is often just moving from one URL to another. It is our intuition that major search engines like Google, Yahoo and MSN Live, as members of what we call the Web Infrastructure (WI), likely have crawled the content and possibly even stored a copy in their cache. Therefore the content is not lost, it "just" needs to be rediscovered. The WI, explored in detail in [21, 17, 11], also includes (besides search engines) non-profit archives such as the Internet Archive (IA) or the European Archive as well as large-scale academic digital data preservation projects e.g., CiteSeer and NSDL.

It is commonplace for content to "move" to different URIs over time. Figure 1 shows two snapshots as an example of a web page whose content has moved within a two year period. Figure 1(a) shows the content of the original URL of the Hypertext 2006 conference[1] as displayed in 1/2009. The original URL clearly does not hold conference related content anymore. Our suspicion is that the website administrators did not renew the domain registration and therefore someone else took over. However, the content is not lost. Figure 1(b) shows the content which is now available at a new URL[2]. This example describes the retrieval problem we

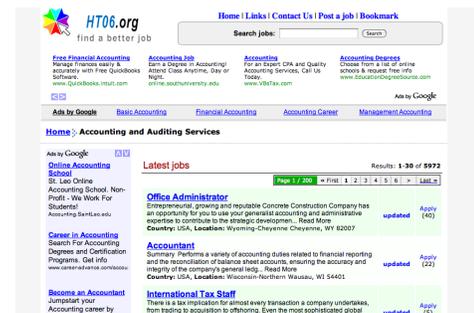

(a) Original URL, new (unrelated) Content

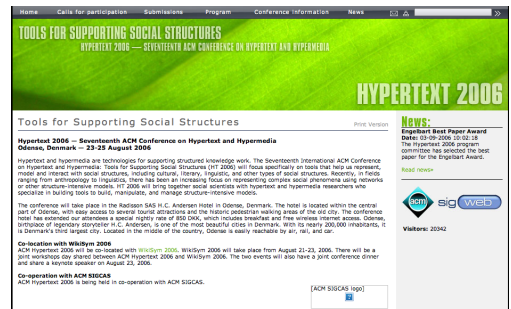

(b) Original Content, new URL

**Figure 1: The Content of the Website for the Conference Hypertext 2006 has Moved over Time**

are addressing with our research. In Figure 2 we are displaying our scenario for discovering web pages that are considered missing. The occurrence of an 404 error is displayed in the first step. Note that a page returning unrelated content (such as in the example above) can be considered missing as well since the user intents to retrieve the original content. Search engine caches and the IA will consequently be queried with the URL requested by the user. In case older copies of the page are available they can be offered to the user. If the user's information need is satisfied, nothing further needs to be done (step (2)). If this is not the case we need to proceed to step (3) where we extract titles, try to obtain tags about the URL and generate LSs from the obtained copies. They are then queried against live search engines and the returned results are again offered to the user as depicted in step (4) of Figure 2. In case the user is again not pleased with the outcome more sophisticated and complex methods need to be applied (step (5)). For example, search engines can be queried to discover pages linking to the missing page. The assumption is that the aggregate of those pages is likely to be about the same topic. From this link neighborhood a LS can be generated. At this point the approach is the same as the LS method, with the exception that the LS has been generated from a link neighborhood and not a cached copy of the page itself. This scenario also needs to be applied in case no copies of the missing page can be found in search engine caches and the IA. The final results are provided in step (6). The important point of this scenario is that it works while the user is browsing and therefore has to provide results in real time.

---
[1] http://www.ht06.org/

[2] http://hypertext.expositus.com/

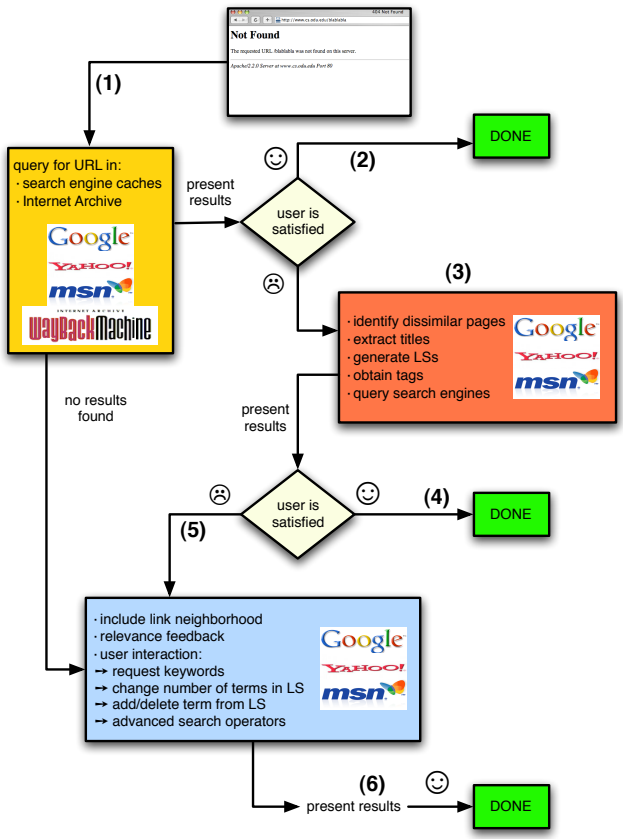

Figure 2: Process to Rediscover Missing Web Pages

Recent research has shown that lexical signatures (LSs) generated from the textual content of web pages are suitable as search engine queries to rediscover missing pages [22, 13]. LSs are rather expensive to generate, the web pages' titles however are available at a lower cost. We investigated the change of web page content compressed into LSs over time in [13] and focus here on the issue of title changes over time. Our intuition is that if the frequency of change is high, titles may not be very useful after all for rediscovering a missing web page. In this paper we present the preliminary results of a study investigating the frequency and degree of change of web pages' titles over time. We predict a lower degree of change compared to LSs since LSs are based on the content of the entire page which supposedly changes more frequently than the general topic captured by the page title. The Appendix shows three examples of web pages, their titles as observed over time by the IA and our computed similarity scores.

## 2. RELATED WORK

### 2.1 Missing Web Pages

Missing web pages are a pervasive part of the web experience. The lack of link integrity on the web has been addressed by numerous researchers [5, 6, 1, 2]. In 1997 Brewster Kahle published an article focused on preservation of Internet resources claiming that the expected lifetime of a web page is 44 days [12]. A different study of web page availability performed by Koehler [14] shows the random test collection of URLs eventually reached a "steady state" after approximately 67% of the URLs were lost over a 4-year period. Koehler estimated the half-life of a random web page is approximately two years. Lawrence et al. [15] found in 2000 that between 23 and 53% of all URLs occurring in computer science related papers authored between 1994 and 1999 were invalid. By conducting a partially manual search on the Internet, they were able to reduce the number of inaccessible URLs to 3%. This confirms our intuition that information is rarely lost, it is just moved. This intuition is also supported by Baeza-Yates et al. [3] who show that a significant portion of the web is created based on already existing content.

Spinellis [24] conducted a study investigating the accessibility of URLs occurring in papers published in Communications of the ACM and IEEE Computer Society. He found that 28% of all URLs were unavailable after five years and 41% after seven years. He also found that in 60% of the cases where URLs where not accessible, a 404 error was returned. He estimated the half-life of an URL in such a paper to be four years from the publication date. Dellavalle et al. [7] examined Internet references in articles published in journals with a high impact factor (IF) given by the Institute for Scientific Information (ISI). They found that Internet references occur frequently (in 30% of all articles) and are often inaccessible within months after publication in the highest impact (top 1%) scientific and medical journals. They discovered that the percentage of inactive references (references that return an error message) increased over time from 3.8% after 3 month to 10% after 15 month up to 13% after 27 month. The majority of inactive references they found were in the *.com* domain (46%) and the fewest in the *.org* domain (5%). By manually browsing the IA they were able to recover information for about 50% of all inactive references.

### 2.2 Search Engine Queries

The work done by Henzinger et al. [9] is related in the sense that they tried to determine the "aboutness" of news documentations. They provide the user with web pages related to TV news broadcasts using a 2-term summary which can be thought of as a LS. This summary is extracted from closed captions of the broadcast and various algorithms are used to compute the scores determining the most relevant terms. The terms are used to query a news search engine while the results must contain all of the query terms. The authors found that 1-term queries return results that are too

| Length | URLs/Domain | | | | |
|---|---|---|---|---|---|
| | .com | .org | .net | .edu | total |
| 1 | 392 | 66 | 27 | 13 | 498 |
| 2 | 239 | 64 | 39 | 40 | 382 |
| 3 | 92 | 20 | 8 | 19 | 139 |
| 4 | 35 | 11 | 3 | 12 | 61 |
| 5 | 6 | | | 3 | 9 |
| 6 | 1 | | | | 1 |
| $\sum$ | **765** | **161** | **77** | **87** | **1090** |

Table 1: URL Character Statistics

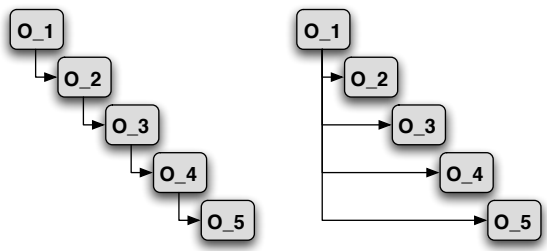

Figure 3: Sliding and Rooted Comparison Methods

vague and 3-term queries return too often zero results. Thus they focus on creating 2-term queries.

He and Ounis' work on query performance prediction [8] is based on the TREC dataset. They measured retrieval performance of queries in terms of average precision (AP) and found that the AP values depend heavily on the type of the query. They further found that what they call *simplified clarity score (SCS)* has the strongest correlation with AP for title queries (using the title of the TREC topics). SCS depends on the actual query length but also on global knowledge about the corpus such as document frequency and total number of tokens in the corpus.

### 2.3 The Web Infrastructure for the Preservation of Web Pages

Nelson et al. [21] present various models for the preservation of web pages based on the web infrastructure. They argue that conventional approaches to digital preservation such as storing digital data in archives and applying methods of refreshing and migration are, due to the implied costs, unsuitable for web scale preservation.

McCown has done extensive research on the usability of the web infrastructure for reconstructing missing websites [17]. He also developed *Warrick* [19], a system that crawls web repositories such as search engine caches (characterized in [18]) and the index of the IA to reconstruct websites. His system is targeted to individuals and small scale communities that are not involved in large scale preservation projects and suffer the loss of websites.

## 3. EXPERIMENTAL SETUP

### 3.1 Data Gathering

It is the main objective of this experiment is to investigate the (degree of) change of web pages' titles over time. It is clearly unfeasible to download all pages from the web on a regular bases over time and analyze their changes. On the other hand it has been shown that finding a small set of web pages that are representative for the entire web is not trivial [10, 23, 25]. We chose to randomly sample 6,000 URLs from the Open Directory Project at dmoz.org. There is an implicit bias in this selection but it appears more suitable than attempting to get an unbiased sample and therefore for the sake of simplicity it shall be sufficient.

We crawled the 6,000 pages and randomly extracted from each of the pages up to three URLs which are referencing to locations within the same top level domain. The resulting set theoretically contains 18,000 URLs. In practice this number is lower since a number of URLs did not contain any links or were simply inaccessible to the crawler at the time of the crawl in February of 2009. Similar to the filters applied in [22] (also with the implicit bias towards English language web pages) we dismissed URLs that were not from the .com, .net, .org or .edu domains. In order to investigate the temporal change of web page titles we checked the availability of all remaining URLs in the IA and found copies for a total of 1090 URLs. We call one particular copy of a web page in the IA identified by a time stamp an *observation*. We downloaded a total of more than 100,000 observations for our 1090 URLs. Table 1 summarizes the characteristics of all 1090 URLs that have observations in the IA. The length of an URL is the number of tokens the path to the referenced object contains. For example the URLs foo.bar/ and foo.bar/index.html have a length of one and foo.bar/bar/ as well as foo.bar/bar/index.html have a length of two. URLs from the .com domain (70.2%) as well as URLs of length one (45.7%) and two (35%) are dominant in our sample set.

### 3.2 Measures of Change

With the corpus created we analyze the change of web page titles over time with two different measures. Since we anticipate a low degree of change we first investigate the general frequency of change meaning how often a title is modified over the time span covered by all available IA observations.

The second measure is meant to represent the degree of change of the titles over time. We use the Levenshtein score which captures the minimum number of operations needed to transform one title into another and compute it for all titles of our corpus. A low Levenshtein score means the compared titles are very dissimilar and a high score indicates a high level of similarity. The score is different from what is known as the Levenshtein distance where the value of 1.0 means totally dissimilar strings and 0 indicates a match. We compute the score in two different ways: the *sliding* and the *rooted* comparison. To explain the two methods let us consider an URL with five observations $O_1...O_5$. The sliding comparison computes the Levenshtein score between $O_1$ and $O_2$, $O_2$ and $O_3$, $O_3$ and $O_4$ and $O_4$ and $O_5$. It continuously slides the comparison window forward by one observation, hence the name. The rooted method (for the same example) will compute the score between $O_1$ and $O_2$, $O_1$ and $O_3$, $O_1$ and $O_4$ and $O_1$ and $O_5$ hence we call it a rooted comparison. This example is visually represented in Figure 3.

We used the SimMetrics library[3] to compute the Levenshtein scores.

---
[3] http://www.dcs.shef.ac.uk/~sam/simmetrics.html

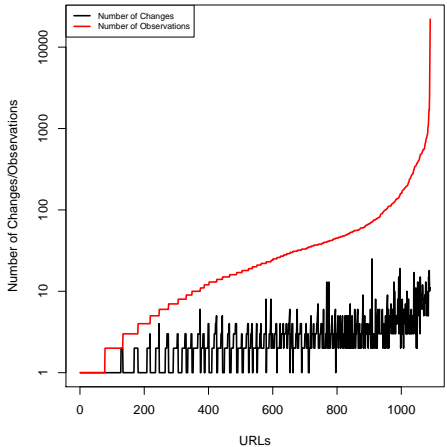

**Figure 4: Number of Title Changes and Observations in the Internet Archive of all URLs**

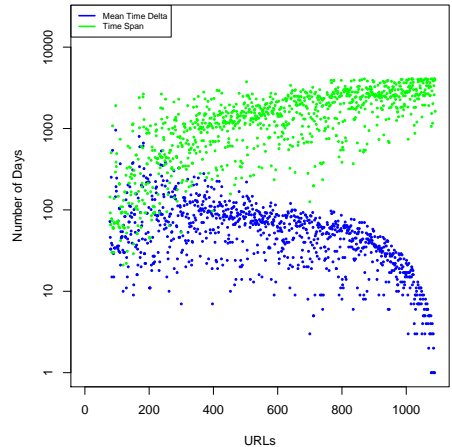

**Figure 5: Mean Time Delta Between all Observations in the Internet Archive and Entire Time Span of Observations (in Days) of all URLs**

## 4. EXPERIMENTAL RESULTS

### 4.1 The Number of Changes

Each time a title changes, that means the captured title of observation $O_{n+1}$ is different compared to the title of the earlier observation $O_n$ the frequency of change is increased by one. Figure 4 shows in semi-log scale the number of title changes and the total number of IA observations (y-axis) of all 1090 URLs (x-axis). The URLs are sorted in increasing order by number of observations first and number of title changes second. We generally observe a rather low frequency of change. The most "inconsistent" URL accounts for 25 title changes. This result confirms the intuition that titles are more stable than for example LSs of web pages. The number of observations of URLs in the IA over time does not impact the number of changes of their titles. For example we see URLs with thousands of observations having similarly few title changes as URLs with less than 50 observations. This means that the frequency of title changes in our sample set is not biased towards the number of available observations in the IA.

Figure 5 is also plotted in semi-log scale. It displays the mean time that has passed between all available IA observations as well as the amount of time passed between the first and the last observation. Both values are measured in days and indicated on the y-axis. The ordering of the URLs in this graph is the same as in Figure 4. We can see that with the increasing number of IA observations per URL the time gap between observation decreases. The overall time span passed between the first and the last observation starts off high and slightly increases with the rising number of IA observations. This result indicates that URLs with many observations in the IA have been crawled frequently in the past in a rather short period of time and most likely are still being crawled with that frequency. It further points to an early start of the crawl for such URLs since the overall time span of all observations is high. Since the web is growing and the IA claims to constantly increase the number of pages crawled ([20]) this observation matches our intuition.

However we are not in the position to say whether just the frequency of crawls for already indexed pages increased or the actual size of the index has increased meaning new pages have been discovered, crawled and indexed. For URLs with 10 or less observations the difference between the two time values is hardly noticeable.

### 4.2 Degree of Change

As mentioned above the degree of change is measured using the Levenshtein score. The score varies between zero and one where one means the titles are identical and zero means they are completely dissimilar. The mean sliding scores over all observations per URL are shown in Figure 6. These scores are generally very high. Only five out of our 1090 URLs have a score of zero and more than 85% of all URLs show a score of 0.8 or above. That means that titles generally do not change drastically between a pair of observations. Slight changes are much more likely. The mean rooted values are plotted in Figure 7 and they are as we anticipated lower. Even though only nine URLs have the zero score just about 56% of the URLs have a score equal or above 0.8. This result confirms that a lot of the titles in our sample set do change compared to their first available IA observation but still not as dramatic as the pages' content for example (see [13]).

## 5. CONCLUSIONS AND FUTURE WORK

First results have confirmed our intuition that web pages' titles are a good resource for rediscovering missing pages. We therefore further investigate the potential of such titles by analyzing their frequency and degree of change over time. We randomly sampled URLs from dmoz.org and analyzed their titles from copies available through the IA. We found a low frequency of change. For example, from all URLs with at least 10 observations (68%) almost 89% show changes in their title only five times or less. We analyzed the degree of change by computing Levenshtein scores for a sliding and rooted comparison of all observations per URL. The scores are very high for the sliding measure. More than five out of six URLs have a score of 0.8 which means the title changes if at all just slightly. The rooted scores are lower but still more than one half of the URLs have a score of 0.8 or above.

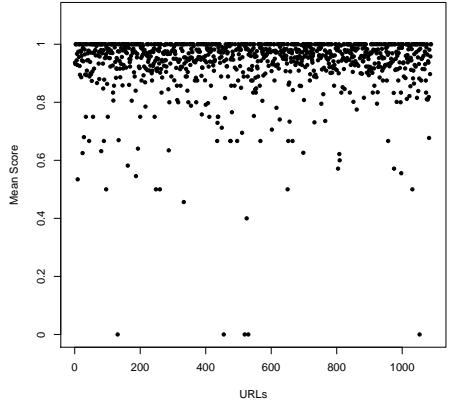

**Figure 6: Mean Levenshtein Score of all Titles - Sliding Comparison**

We consider this work as preliminary and see several aspects for future work. Most importantly we will apply various natural language processing techniques to create a "quality prediction" model for web pages' titles. The goal is to predict how promising any given title is for our purpose in order to decide whether to use the title or maybe rather generate a LS of the page. Since we are using these titles to query search engines (and each query comes with an associated cost) in order to rediscover missing web pages, we will then be able to automatically dismiss low value titles such as *Index* and *Home Page*. Another interesting aspect for future work is investigating the change of titles for dynamic compared to static URLs. We can for example identify URLs that are passing parameters with a *&* as dynamic. However the difficult part is to determine when URIs that do not contain such parameters resolve dynamic content.

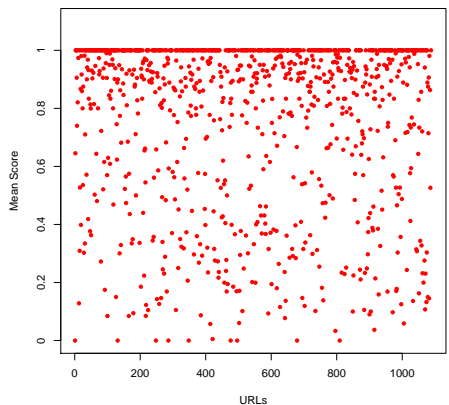

**Figure 7: Mean Levenshtein Score of all Titles - Rooted Comparison**

## 6. REFERENCES

[1] H. Ashman. Electronic document addressing: Dealing with change. *ACM Computing Surveys*, 32(3):201–212, 2000.

[2] H. Ashman, H. Davis, J. Whitehead, and S. Caughey. Missing the 404: Link integrity on the world wide web. In *Proceedings of WWW '98*, pages 761–762, 1998.

[3] R. Baeza-Yates, Álvaro Pereira, and N. Ziviani. Genealogical trees on the web: a search engine user perspective. In *WWW '08: Proceeding of the 17th international conference on World Wide Web*, pages 367–376, 2008.

[4] T. Berners-Lee. Cool URIs don't change, 1998. http://www.w3.org/Provider/Style/URI.html last visited in 01/2009.

[5] H. C. Davis. Hypertext link integrity. *ACM Computing Surveys*, page 28.

[6] H. C. Davis. Referential integrity of links in open hypermedia systems. In *Proceedings of HYPERTEXT '98*, pages 207–216, 1998.

[7] R. P. Dellavalle, E. J. Hester, L. F. Heilig, A. L. Drake, J. W. Kuntzman, M. Graber, and L. M. Schilling. INFORMATION SCIENCE: Going, Going, Gone: Lost Internet References. *Science*, 302(5646):787–788, 2003.

[8] B. He and I. Ounis. Inferring Query Performance Using Pre-retrieval Predictors. In *Proceedings of SPIRE '04*, pages 43–54, 2004.

[9] M. Henzinger, B.-W. Chang, B. Milch, and S. Brin. Query-free News Search. In *Proceedings of WWW '03*, pages 1–10, 2003.

[10] M. R. Henzinger, A. Heydon, M. Mitzenmacher, and M. Najork. On Near-Uniform URL Sampling. *Computer Networks*, 33(1-6):295–308, 2000.

[11] A. Jatowt, Y. Kawai, S. Nakamura, Y. Kidawara, and K. Tanaka. A Browser for Browsing the Past Web. In *Proceedings of WWW '06*, pages 877–878, 2006.

[12] B. Kahle. Preserving the Internet. *Scientific American*, 276:82–83, March 1997.

[13] M. Klein and M. L. Nelson. Revisiting Lexical Signatures to (Re-)Discover Web Pages. In *Proceedings of ECDL '08*, pages 371–382, 2008.

[14] W. C. Koehler. Web Page Change and Persistence - A Four-Year Longitudinal Study. *Journal of the American Society for Information Science and Technology*, 53(2):162–171, 2002.

[15] S. Lawrence, D. M. Pennock, G. W. Flake, R. Krovetz, F. M. Coetzee, E. Glover, F. A. Nielsen, A. Kruger, and C. L. Giles. Persistence of Web References in Scientific Research. *Computer*, 34(2):26–31, 2001.

[16] C. C. Marshall, F. McCown, and M. L. Nelson. Evaluating Personal Archiving Strategies for Internet-based Information. In *Proceedings of IS&T Archiving '07*, pages 48–52, 2007.

[17] F. McCown. *Lazy Preservation: Reconstructing Websites from the Web Infrastructure*. PhD thesis, Old Dominion University, 2007.

[18] F. McCown and M. L. Nelson. Characterization of search engine caches. In *Proceedings of IS&T Archiving 2007*, pages 48–52, May 2007.

[19] F. McCown, J. A. Smith, and M. L. Nelson. Lazy

Preservation: Reconstructing Websites by Crawling the Crawlers. In *Proceedings of WIDM '06*, pages 67–74, 2006.
[20] G. Mohr, M. Stack, I. Ranitovic, D. Avery, and M. Kimpton. Introduction to heritrix, an archival quality web crawler. 4th International Web Archiving Workshop (IWAW04), September 2004.
[21] M. L. Nelson, F. McCown, J. A. Smith, and M. Klein. Using the Web Infrastructure to Preserve Web Pages. *IJDL*, 6(4):327–349, 2007.
[22] S.-T. Park, D. M. Pennock, C. L. Giles, and R. Krovetz. Analysis of Lexical Signatures for Finding Lost or Related Documents. In *Proceedings of SIGIR '02*, pages 11–18, 2002.
[23] P. Rusmevichientong, D. M. Pennock, S. Lawrence, and C. L. Giles. Methods for Sampling Pages Uniformly from the World Wide Web. In *AAAI Fall Symposium on Using Uncertainty Within Computation*, pages 121–128, 2001.
[24] D. Spinellis. The decay and failures of web references. *Communications of the ACM*, 46(1):71–77, 2003.
[25] M. Theall. Methodologies for Crawler Based Web Surveys. *Internet Research: Electronic Networking and Applications*, 12:124–138, 2002.

# Appendices
## www.originalbristol.com
mean Levenshtein score sliding: 0.81 rooted: 0.47

```
2007-03-31
Original 106.5 - Bristol

2007-04-17
Original Bristol 106.5 fm Weblog

2007-04-22
Original Bristol 106.5 fm prelaunch blog

2007-05-18
Original Bristol

2007-08-16
Original 106.5 fm - The new radio
station for Bristol - Original like you

2007-12-01
Original 106.5 - Bristol's Best Music!

2008-01-04
Original 106.5
```

## www.sun.com/solutions
mean Levenshtein score sliding: 0.84 rooted: 0.29

```
1998-01-27
Sun Software Products Selector Guides -
Solutions Tree

1999-02-20
Sun Software Solutions

2002-02-01
Sun Microsystems Products

2002-06-01
Sun Microsystems - Business & Industry
Solutions

2003-08-01
Sun Microsystems - Industry &
Infrastructure Solutions

2004-02-02
Sun Microsystems - Solutions

2004-06-10
Gateway Page - Sun Solutions

2006-01-09
Sun Microsystems Solutions & Services

2007-01-03
Services & Solutions

2007-02-07
Sun Services & Solutions

2008-01-19
Sun Solutions
```

## www.datacity.com/mainf.html
mean Levenshtein score sliding: 0.68 rooted: 0.15

```
2000-06-19
DataCity of Manassas Park Main Page

2000-10-12
DataCity of Manassas Park sells Custom
Built Computers & Removable Hard Drives

2001-08-21
DataCity a computer company in Manassas
Park sells Custom Built Computers & Removable
Hard Drives

2002-10-16
computer company in Manassas Virginia sells
Custom Built Computers with Removable Hard
Drives Kits and Iomega 2GB Jaz Drives
(jazz drives) October 2002 DataCity
800-326-5051 toll free

2006-03-14
Est 1989 Computer company in Stafford
Virginia sells Custom Built Secure
Computers with DoD 5200.1-R Approved
Removable Hard Drives, Hard Drive Kits
and Iomega 2GB Jaz Drives (jazz drives),
introduces the IllumiNite® lighted
keyboard DataCity 800-326-5051 Service
Disabled Veteran Owned Business SDVOB
```